\documentclass[amsmath, pra, reprint, superscriptaddress, showpacs,aps, 10pt]{revtex4-1}

\usepackage{graphicx}
\usepackage{dcolumn}
\usepackage{bm}
\usepackage{wasysym}
\usepackage{upgreek}

\usepackage{color}

\begin{document}


\title{Optically active mechanical modes of tapered optical fibers}
\author{C. Wuttke}
\affiliation{Vienna Center for Quantum Science and Technology, TU Wien -- Atominstitut, 1020 Wien, Austria}
\author{G.~D. Cole}%
\affiliation{Vienna Center for Quantum Science and Technology, University of Vienna -- Faculty of Physics, 1090 Wien, Austria}
\author{A. Rauschenbeutel}%
\email{Arno.Rauschenbeutel@ati.ac.at.}
\affiliation{Vienna Center for Quantum Science and Technology, TU Wien -- Atominstitut, 1020 Wien, Austria}
\date{\today}

\begin{abstract}
Tapered optical fibers with a nanofiber waist are widely used tools for efficient coupling of light to photonic devices or quantum emitters via the nanofiber's evanescent field. In order to ensure well-controlled coupling, the phase and polarization of the nanofiber guided light field have to be stable.  Here, we show that in typical tapered optical fibers these quantities exhibit high-frequency thermal fluctuations. They originate from high-Q torsional oscillations that opto-mechanically couple to the nanofiber-guided light. We present a simple ab-initio theoretical model that quantitatively explains the torsional mode spectrum and that can be used to design tapered optical fibers with tailored mechanical properties.
\end{abstract}
\maketitle
Tapered optical fibers with a subwavelength-diameter waist (TOFs) feature a strong evanescent field in the waist region and are widely used to efficiently interface light and matter or to couple light into photonic devices, such as micro-resonators or photonic crystals. These applications rely on the stability of the phase and polarization of the nanofiber guided light field as well as on the position of the nanofiber. Achieving this stability is all the more challenging in a high vacuum environment where the mechanical damping due to the surrounding gas is negligible. Here, we experimentally demonstrate that under these conditions torsional mechanical modes exhibit surprisingly high quality factors and lead to resonantly enhanced vibrations that modulate the phase and polarization of the optical mode via the strain-optic effect. Based on an analytic model as well as on experimental measurements, we show that the commonly used exponential radius profile confines a subset of the torsional mechanical modes to the nanofiber section, leading to the high Q-factors. 

From a mechanical point of view, a TOF is a slender cylinder with varying cross-section. Torsional waves in such a structures can be described by a one-dimensional wave equation which has been treated comprehensively in the literature \citep{Pyle1963,Engan1988}. In case of the mechanical modes considered here, their wavelength is much larger than the largest cross-section of the TOF so that no higher order angular and radial modes exist (slender rod approximation). Furthermore, the radius variations occurring in the TOF profile are sufficiently shallow to assume plane wavefronts that are perpendicular to the fiber axis. Under these conditions, the torsional motion can be described by a Webster-type wave equation
\begin{equation}
	\frac{1}{c_t^{2}}\,\partial_t^2\,\phi(t,z)-\partial_z^2\,\phi(t,z) - \left(\frac{\partial_z I_p(z)}{I_p(z)}\right)\,\partial_z \phi(t,z)= 0 \label{eq:wave}
\end{equation}
where $\phi(t,z)$ is the angular displacement amplitude as function of time $t$ and axial position along the fiber $z$ and $\partial_x=\partial/\partial x$ with $x=t,z$. The torsional wave velocity is radius independent and given by $c_t=\sqrt{G/\rho}=(3680\pm130)~\mathrm{m/s}$, where $G=(30\pm2)$~GPa is the shear modulus and  $\rho=(2210\pm10)~\mathrm{kg/m^3}$ is the mass density of silica \citep{Spinner1962,Borrelli1968,Pfaender1983}. The third term of the wave equation takes the radius profile of the cylinder, $a(z)$, into account via the polar angular moment of inertia $I_p(z)=\pi\,\rho\,a(z)^4/2$. 

\begin{figure}
\includegraphics[width=8.4cm]{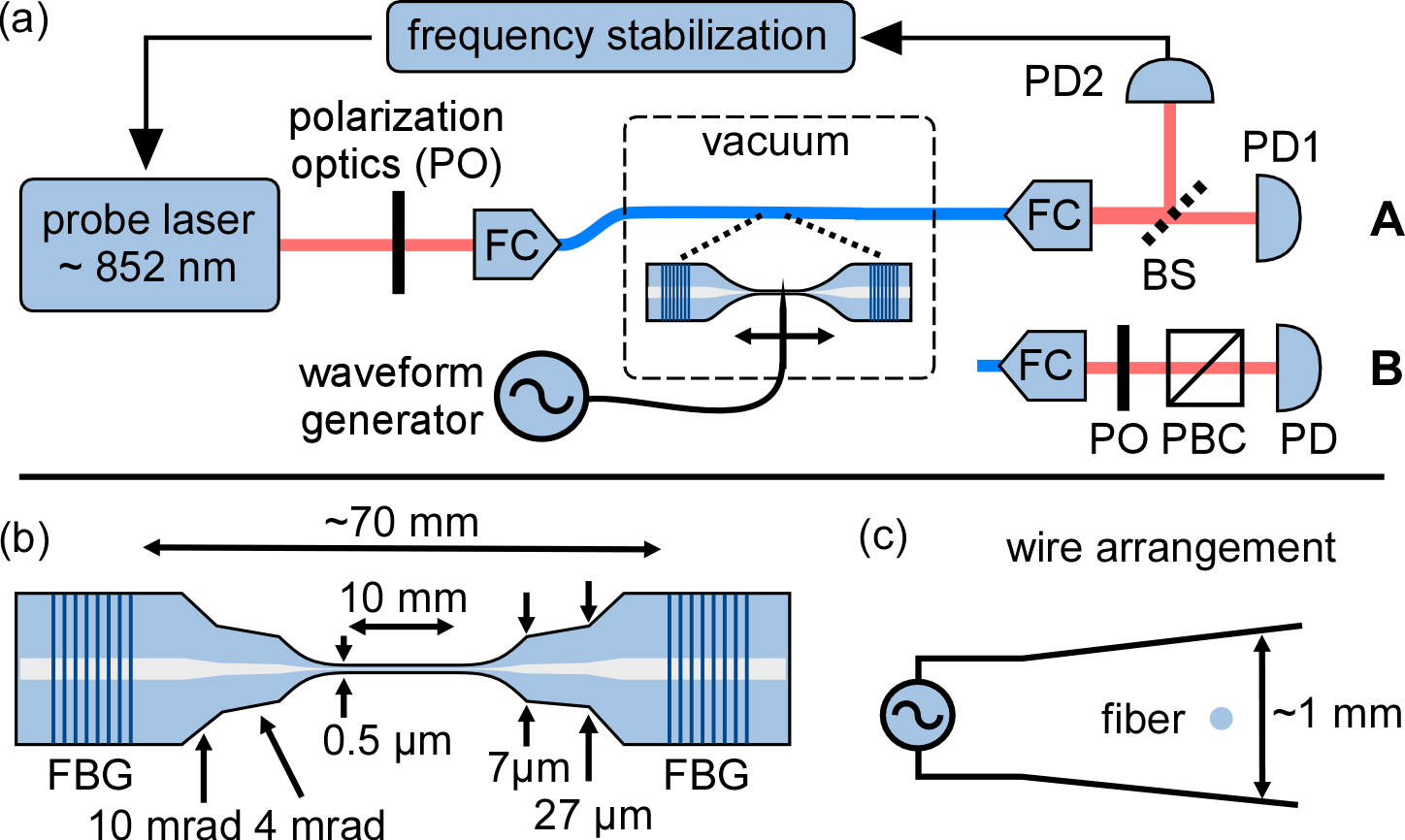}
\caption{(a) Schematical view of the experimental setup, see text for details. (b) Schematic cross-section profile of the TOF-resonator where the radii at the section boundaries and the full opening angles of the conical sections are given. (c) Schematical view of the electrode arrangement used to excite the mechanical modes that consists of two wires that are placed above and blow of the nanofiber. The latter is oriented prependicular to the figure.}\label{fig:setup}
\end{figure}

Using a separation ansatz, one finds $\phi(t,z) = \Phi(z)\,\cos(\omega\,t+\theta)$, where $\omega$ is the angular frequency of the mechanical motion and the phase $\theta$ is fixed by the initial conditions. The remaining differential equation for $\Phi(z)$ can then be solved for the cylindrical, conical, and exponential sections of which the TOF is composed, see Fig.~\ref{fig:setup}(b). Last, the amplitude and torque are matched at the interfaces between these sections. The conical radius profile is given by $a(z)=\kappa\,z$ for which we obtain a Bessel-type differential equation with the solutions \citep{Pyle1963} $z^{-3/2}J_{\pm3/2}(k_0 z)$ where $J_{n}$ is the Bessel function of the first kind of order $n$ and $k_0=\omega/c_t$. In the case of the exponential horn, where $a(z)=a_0 \exp(\alpha\,z/2)$, Eq.~\ref{eq:wave} becomes:
\begin{equation}
	\partial_z^2\,\Phi(z) + 2\alpha\,\partial_z \Phi(z)+k_0^2\,\Phi(z)= 0~. \label{eq:wave:exphorn}
\end{equation}
This differential equation yields a cut-off frequency $\omega^{co}_{t}=\alpha\,c_t$ below which $\Phi(z)$ exhibits no nodes (see Appendix). In this case, $\Phi(z)$ decreases exponentially along $z$, thereby imposing a strict requirement on the mode shape of the neighboring TOF sections. For higher frequencies than $\omega^{co}_{t}$, however, $\Phi(z)$ has nodes in the exponential horn and extends further into latter. 

We briefly mention two important modifications of $c_t$ which typically occur in nanofibers: Even small axial forces create high axial strain $\epsilon$ in nanofibers which modifies $c_t$ by a factor \citep{Chi1984} of $(1+2(1+\nu_p)\,\epsilon)^{1/2}$, where  $\nu_p=0.168$ is Poisson's ratio \citep{Borrelli1968}. Close to the rupture point of the nanofiber \citep{Brambilla2009}, one obtains an increase of $c_t$ by $18~\%$. In some experiments, high optical powers are used which can heat the nanofiber up to temperatures of 1000~K and beyond \citep{Vetsch2010,Goban2012,Wuttke2013}. The value of Young's modulus and Poisson's coefficient increase from room temperature to $1400$~K by $11~\%$ \citep{Spinner1962}, thereby increasing $c_t$ by up to $5~\%$.

The TOFs are fabricated from commercial fused silica optical fibers with a cladding diameter of $125~\upmu$m in a heat and pull process \citep{Warken2008} which allows us to fabricate TOFs of predetermined shape with a relative radius uncertainty of $\pm 10$~\%~ \citep{Stiebeiner2010} and a relative radius homogeneity in the waist region of $1~\%$ (peak to peak) \citep{Wiedemann2010}. The TOF radius profile is schematically shown in Fig.~\ref{fig:setup}(b) and consists of a cylindrical nanofiber waist with a radius of $r_w=500$~nm and a length of $l_w=10$~mm which widens in two taper sections on both sides to the initial fiber diameter. Each of the latter is composed of two conical sections with different opening angles which are connected to the waist by a $9$~mm long section with an exponential radius profile where $\alpha = 0.76~\mathrm{mm^{-1}}$ (see above). After the fabrication process, the fibers are strained by $\epsilon_w \approx2~\%$ of the nominal waist length and glued to an U-shaped aluminum holder. 

The measurement scheme used here detects periodic refractive index variations in the material that are caused by the mechanical vibrations via the strain-optic effect  \citep{Vedam1950, Borrelli1968}. They are observed via polarization fluctuations of the transmitted light or via optical path length variations using a fiber-integrated optical Fabry-P\'erot type cavity  \citep{Wuttke2012}. The latter consists of two fiber Bragg grating mirrors (FBGs) with a stop-band at a central wavelength of $\lambda_\mathrm{FBG}=852$~nm and a spectral width of $0.2$~nm that are located on both sides of the TOF in the untapered portion of the fiber, see Fig.~\ref{fig:setup}(b). We probe the resonator using  laser light with a wavelength within the FBG stop-band that is launched into the TOF via a fiber coupler (FC), see Fig.~\ref{fig:setup}(a). The TOF is placed inside a vacuum chamber with a minimal pressure of $10^{-7}$~mbar in order to minimize the mechanical damping from the background gas. At the output port of the fiber, the transmitted light is split by a 50/50 beam splitter (BS) and detected with two photodiodes (PD1 and PD2). The signal of PD2 is used to stabilize the frequency of the laser to the side of a Fabry-P\'erot fringe at half maximum, thereby maximizing the sensitivity to optical path length changes and the dynamic range. The stabilization bandwidth of 2~kHz is chosen much smaller than the frequencies of the mechanical modes under study.

For the method in which the polarization fluctuations are detected, we set the laser to a wavelength outside the FBG stop-band or, alternatively, use a TOF without a resonator. A polarizing beam splitter cube (PBC) converts the polarization fluctuations to intensity variations that are detected by a photodiode at the PBC's transmission port. Polarization optics (PO) before and after the fiber are used to optimize the signal strength. In a addition to passive thermal driving, the mechanical modes can be actively excited by an electric field that exerts an alternating force on stray charges on the insulating silica surface. This field is localized along the nanofiber axis and is generated by two thin and movable electrodes placed above and below the fiber, see Fig.~\ref{fig:setup}(a,c). 

\begin{figure}
	\centering
	\includegraphics[width=8.4cm]{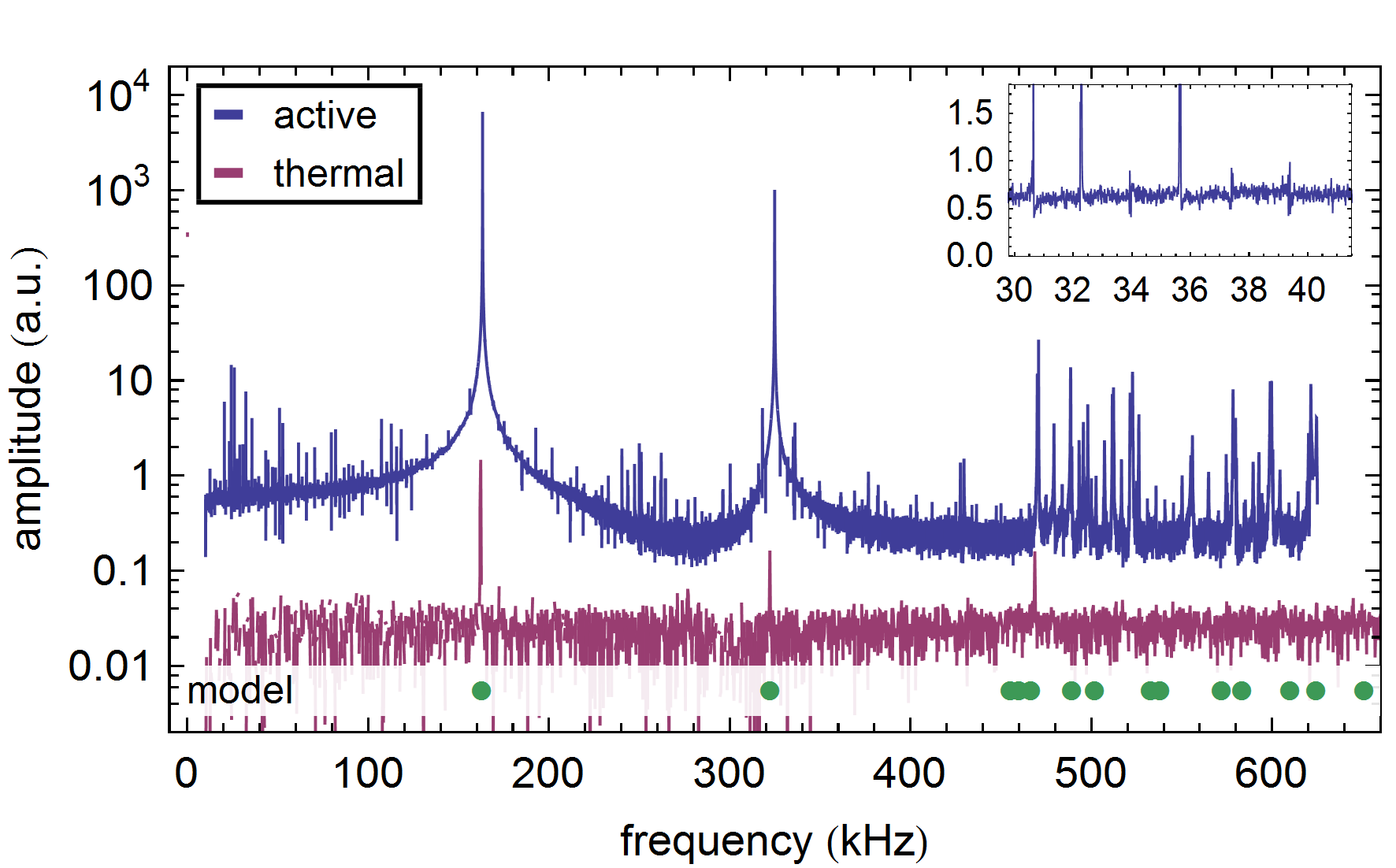}
	\caption{Spectral amplitude detected from polarization fluctuations of two samples with the same radius profile, measured with thermal (lower purple line) and external excitation (upper blue line). The green markers show the model prediction. Inset: Magnified view of the response function in the spectral range around 35~kHz\label{fig:spectrum}}
\end{figure}

We first measure the amplitude spectrum of the thermal motion using the polarization method. The resulting spectrum is shown as a purple line in Fig.~\ref{fig:spectrum}.
The spectrum features three peaks at frequencies of $162$~kHz, $321$~kHz, and $469$~kHz, corresponding to three mechanical resonances. These frequencies are approximately integer multiples which indicates that they are harmonics of the same mechanical mode and that the lowest frequency stems from the fundamental mode. We compare this measurement to the spectral response of a TOF with the same radius profile subjected to electric field excitation at a fixed position. For this purpose, a sinusoidal voltage modulation is applied to the electrodes while recording the amplitude at the modulation frequency using a lock-in detection scheme. We obtain a spectral response which shows resonances that coincide with previously observed ones, see blue line in Fig.~\ref{fig:spectrum}. This confirms the reproducibility of our fabrication method. Additionally, we find resonances with a much smaller frequency spacing in the frequency range beyond 470~kHz. This frequency matches the theoretical cut-off frequency of $\omega^{co}_t=2\pi\,450$~kHz for the exponential taper sections. The higher spectral mode density stems from the larger spatial extent of the modes above the cut-off frequency. Furthermore, we observe resonances with a frequency spacing of $\approx 1.8$~kHz over the full spectral range, see Fig.~\ref{fig:spectrum} inset. By additional finite element method calculations, we confirm that they correspond to transversal modes. In contrast to the torsional modes, transversal modes are not confined by the taper but extend over the full length of the TOF, thereby resulting in the small frequency spacing. 

We make an ab-initio theory prediction for the torsional resonance frequencies that is derived from the design radius profile of the TOF and the bulk elastic constants of silica. We omit modes which exhibit a high amplitude near the clamping points and which are therefore expected to suffer from high clamping losses. Such modes stem from the torsional oscillation of the two ends of the TOF which are weakly coupled by the nanofiber. In the simulations, they can be clearly identified via their spatial mode profile and because they come in pairs with a small frequency splitting between the odd and even mode. The predicted frequencies for the remaining modes are marked in Fig.~\ref{fig:spectrum} by green disks and show excellent agreement with the observed fundamental and second harmonic torsional resonance frequencies. Furthermore, the theory reproduces the close mode spacing above the cut-off frequency. The precise resonance frequencies in this frequency range are not correctly predicted by the theory. We attribute this fact to deviations between the assumed and the actual radius profile. 

\begin{figure}
	\centering
	\includegraphics[width=8.4cm]{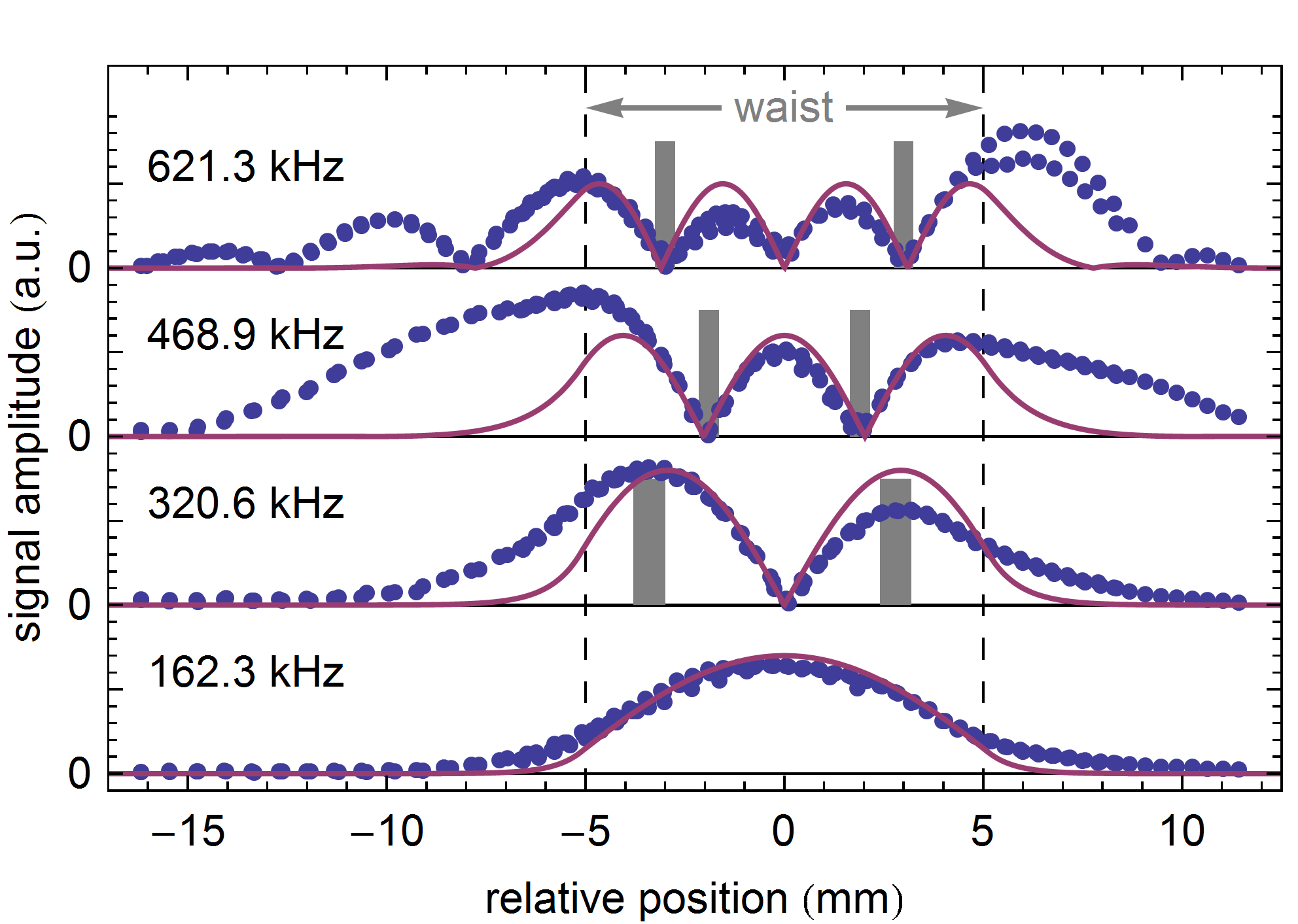}
	\caption{Absolute value of the signal amplitude as function of the electrode position for selected modes (circles). The waist region is determined from the the symmetry point of the modes and is indicated by dashed lines. The theory prediction of the mode profiles is shown as a purple line for the modes at the frequencies: $\{163.2;322.9;466.8;610.4\}$~kHz.
	\label{fig:mech:modeshape}}
\end{figure}

We now measure the modal amplitude profiles along the TOF axis for selected modes. This is achieved by recording the response of the TOF to fixed-frequency excitation at different axial positions along the fiber, see Fig.~\ref{fig:mech:modeshape}.
We expect that the effective excitation strength depends on the fiber radius. As a consequence, the measured response is only proportional to the modal amplitude profile in the waist region where the radius is constant. The modal amplitude profiles of the resonances at $162$~kHz and $321$~kHz resemble that of the first two harmonics of a cylinder with clamped ends. The two higher modes, however, extend much further into the taper region and the mode at $f=621.3$~kHz exhibits nodes in the exponential section. Hence, we directly observe the effect of the exponential horn: At frequencies below $\omega^{co}_t$, the modes are required to decay exponentially along the taper with a decay constant that becomes more shallow with increasing frequencies. Beyond $\omega^{co}_t$, the modes expand further into the taper with nodes along the fiber axis. We find a good qualitative agreement with the theoretical prediction of the absolute modal amplitude profile, see purple lines in Fig.~\ref{fig:mech:modeshape}.

We now determine the phase velocity of the modes from their frequency and the wavelength using $c_{t} = \lambda\,f$~. The wavelength is inferred from the separation of two nodes or anti-nodes in the waist region. Their position and the corresponding uncertainty are indicated by the center and the width of the gray bars in Fig.~\ref{fig:mech:modeshape}, respectively. We obtain $\lambda/2=(6.2\pm0.4)$~mm at $f=(320.6\pm4.1)$~kHz, $\lambda/2=(3.8\pm0.3)$~mm at $f=(468.9\pm1.9)$~kHz, and $\lambda=(6.0\pm0.3)$~mm at $f=(621.3\pm0.2)$~kHz. We compute $c_t$ using the weighed average of all measurements and obtain
\begin{equation}
	c_{t} = (3737\pm177)\mathrm{~m/s}~.
\end{equation}
This value is in excellent agreement with the literature value of the torsional acoustic velocity in silica \citep{Spinner1962,Borrelli1968,Pfaender1983} $(3680\pm130)\mathrm{~m/s}$ and is clearly different from the ones for axial $(5630\pm170)~\mathrm{m/s}$ and string modes $(1130\pm35)~\mathrm{m/s}$. This results further substantiates the identification of the mechanical modes as torsional modes of the TOF.

We now measure the Q-factors of the individual mechanical modes using a ring-down technique. We determine the amplitude decay time from an exponential fit which we use to compute the Q-factors listed in Tab.~\ref{tab:Q:factors}.
\begin{table}[htb]
	\begin{tabular}{|c|c|c|c|c|c|}
	 \hline
		f (kHz) & Q ($10^4$)& f (kHz) & Q ($10^4$)&f (kHz) & Q ($10^4$)\\ \hline
		148 & $2.60\pm0.05$ &	500 & $0.18\pm0.05$ &	556 & $0.33\pm0.02$ \\ \hline
		302 & $2.30\pm0.05$ &	502 & $0.71\pm0.03$ &	574 & $0.56\pm0.03$	\\ \hline
		401 & $0.18\pm0.07$ &	526 & $0.75\pm0.03$ &	778 & $0.31\pm0.02$	\\ \hline
		441 & $1.03\pm0.03$ &	529 & $0.11\pm0.02$ &  	817 & $0.12\pm0.01$	\\ \hline
		453 & $0.17\pm0.03$ &	545 & $0.40\pm0.03$  & & \\ \hline
	\end{tabular}
	\caption{Q-factors of selected mechanical modes. \label{tab:Q:factors}}
\end{table}
We find high values of $Q\approx2.5\cdot10^{4}$ for the first two harmonics. The mechanical loss in silica has been studied extensively in literature \citep{Pohl2002,Penn2006}. For small structures, such as nanofibers, it is dominated by surface effects which scale with the surface to volume ratio (SVR). Based on this model, we obtain a prediction for the nanofiber waist of $Q=(1.5\pm0.5)\cdot10^{4}$, slightly lower than the measured values. The deviation might be partly due to the fact that the mode extends into the taper region where the SVR is more favorable. Note that the resonance frequencies of the TOF used for this measurement are $9~\%$ smaller than for the previous samples. This is due to a lower mechanical strain applied to the sample.

\begin{figure}
	\centering
	\includegraphics[width=6cm]{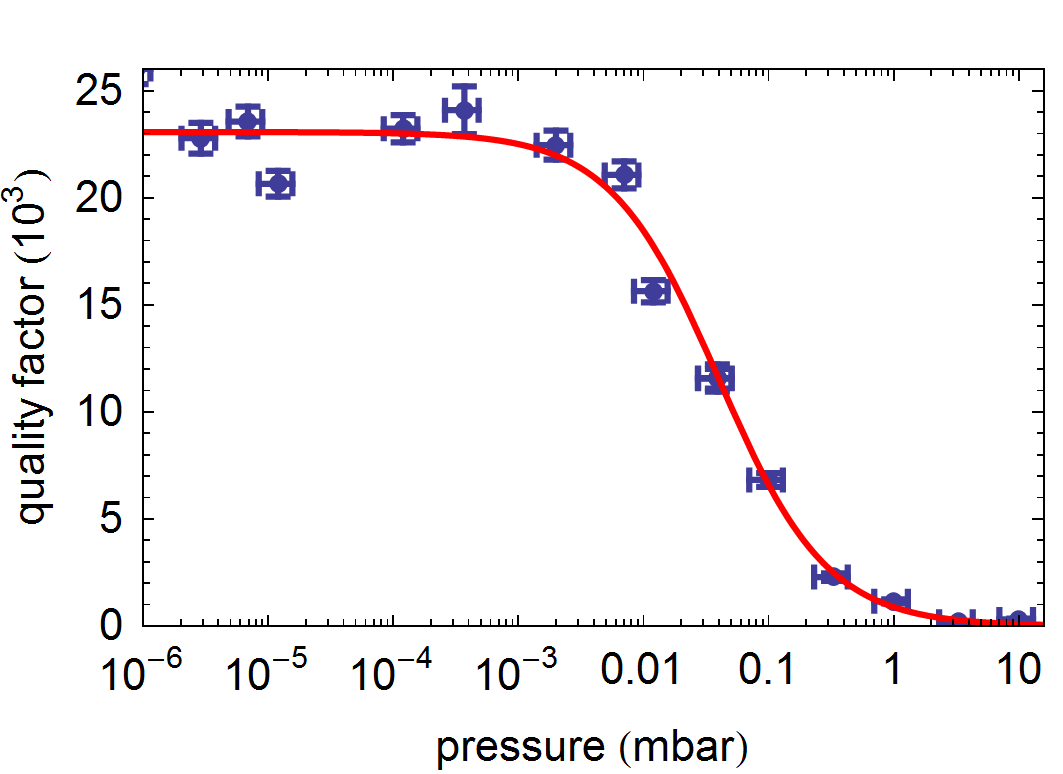}
	\caption{Quality factor of the fundamental torsional mode ($148.1$~kHz) as function of the background gas pressure. Red line: Empirical model fit of the form $Q^{-1}(p)=Q_\mathrm{i}^{-1}+(\eta/p)^{-1}$, where $Q_\mathrm{i}=(2.31\pm 0.06)\cdot 10^{4}$ and $\eta=(923\pm 89)$~mbar. \label{fig:Q:pressure}}
\end{figure}

We also study the damping of the torsional modes by the background gas. For this purpose, we measure the Q-factor of the fundamental mode at different gas pressures, see Fig.~\ref{fig:Q:pressure}.
The Q-factor is approximately constant up to pressures of $p\sim 10^{-3}$~mbar, meaning that the intrinsic losses dominate, and then decreases due to damping by the background gas. We fit the data by an empirical model that assumes that the damping is proportional to the particle density and find excellent agreement, see red line in Fig.~\ref{fig:Q:pressure}.

Summarizing, we show that tapered optical fibers exhibit optically active mechanical modes. The thermal excitation of these modes at room temperature leads to high-frequency fluctuations of the phase and polarization of the nanofiber-guided light. We measured the spectrum of these fluctuations and found resonances at frequencies of several 100~kHz which we identified as torsional modes. We showed that the widely used exponential radius profile confines these modes to the nanofiber section below a cut-off frequency, thereby enabling high Q-factors at low gas pressures. Our analytic ab-initio mechanical model of the tapered optical fibers reproduces the measured spectrum and amplitude profiles. It should thus allow one to design the mechanical spectrum according to experimental needs. 

Our findings have important consequences for the use of tapered optical fibers in a high vacuum environment. As an example, first estimations show that the thermal polarization fluctuations found in this work currently limit the storage time of nanofiber-based atom traps \citep{Vetsch2010,Goban2012}. They might also limit the ultimate ideality of tapered fiber coupling to photonic devices \citep{Spillane2003}. More generally, high-Q torsional resonances may also occur in other kinds of free-standing subwavelength-diameter waveguides and may influence the stability of the guided light fields. This applies in particular to the case of polarization maintaining structures where the torsional vibration directly translates into the polarization rotation. 

\begin{acknowledgements}
We thank Manfred Rothhardt and co-workers for the FBGs and Sam Dawkins, Peter Eiswirt, and Christian Wagner for their contributions to the initial stages of the experiment. We acknowledge financial support by the Volkswagen Foundation (Lichtenberg Professorship), and NanoSci-E+ (NOIs project) and the Austrian Science Fund (FWF; SFB FoQus Project No. F4017).
\end{acknowledgements}

\section*{Appendix}
The spatial part of the wave equation for the torsional wave in an exponential horn, where $a(z)=a_0 \exp(\alpha\,z/2)$, reads:
\begin{equation}
	\partial_z^2\,\Phi(z) + 2\alpha\,\partial_z \Phi(z)+k_0^2\,\Phi(z)= 0~. \label{eq:wave:exphorn}
\end{equation}
The expression above equals the differential equation of the one-dimensional damped harmonic oscillator (DHO) where the spatial coordinate here corresponds to time for the DHO:
\begin{equation}
	\partial_t^2\,y(t,z) + 2\zeta\,\partial_t y(t,z)+\omega_\mathrm{DHO}^2\,y(t,z)= 0~, \label{eq:DHO}
\end{equation}
where $y$ is the displacement, $\omega_\mathrm{DHO}$ is the undamped eigenfrequency and $\zeta$ is the damping constant \citep{Demtroeder1}. Here, the damping regime is determined by the ratio $\zeta/\omega_\mathrm{DHO}$. If the ratio is smaller than one, the DHO performs damped oscillations (weak damping). However, if the ratio is larger than one, no oscillatory solutions exist and the amplitude decreases exponentially (strong damping). The special case of $\zeta/\omega_\mathrm{DHO}=1$ is called critical damping and marks the transition between the two damping regimes. 

From this comparison, we find that the transition between modes that decay exponentially in the exponential horn and those with nodes occurs at $\omega^{co}_{t}=\alpha\,c_t$, the cut-off frequency. This cut-off behavior of the exponential horn controls the mechanical mode-structure of the TOF: For frequencies below $\omega^{co}_{t}$ no solution of propagating waves exists, meaning that no nodes occur within the horn and the amplitude decreases exponentially along $z$. This imposes a strict requirement on the mode at the boundaries to the neighboring TOF sections. Taking the solutions in the cylindrical nanofiber section into account, i.e. the harmonic functions, it can only be met by modes that resemble those of a cylinder with fixed ends and approximately the nanofiber length. Beyond $\omega^{co}_{t}$, however, oscillatory solutions (nodes in the horn) can occur. As a consequence, positive slopes are allowed at the boundaries which results in a larger variety of modal shapes that extend into the exponential sections and a correspondingly smaller mode spacing.


\begin{thebibliography}{19}%
\makeatletter
\providecommand \@ifxundefined [1]{%
 \@ifx{#1\undefined}
}%
\providecommand \@ifnum [1]{%
 \ifnum #1\expandafter \@firstoftwo
 \else \expandafter \@secondoftwo
 \fi
}%
\providecommand \@ifx [1]{%
 \ifx #1\expandafter \@firstoftwo
 \else \expandafter \@secondoftwo
 \fi
}%
\providecommand \natexlab [1]{#1}%
\providecommand \enquote  [1]{``#1''}%
\providecommand \bibnamefont  [1]{#1}%
\providecommand \bibfnamefont [1]{#1}%
\providecommand \citenamefont [1]{#1}%
\providecommand \href@noop [0]{\@secondoftwo}%
\providecommand \href [0]{\begingroup \@sanitize@url \@href}%
\providecommand \@href[1]{\@@startlink{#1}\@@href}%
\providecommand \@@href[1]{\endgroup#1\@@endlink}%
\providecommand \@sanitize@url [0]{\catcode `\\12\catcode `\$12\catcode
  `\&12\catcode `\#12\catcode `\^12\catcode `\_12\catcode `\%12\relax}%
\providecommand \@@startlink[1]{}%
\providecommand \@@endlink[0]{}%
\providecommand \url  [0]{\begingroup\@sanitize@url \@url }%
\providecommand \@url [1]{\endgroup\@href {#1}{\urlprefix }}%
\providecommand \urlprefix  [0]{URL }%
\providecommand \Eprint [0]{\href }%
\providecommand \doibase [0]{http://dx.doi.org/}%
\providecommand \selectlanguage [0]{\@gobble}%
\providecommand \bibinfo  [0]{\@secondoftwo}%
\providecommand \bibfield  [0]{\@secondoftwo}%
\providecommand \translation [1]{[#1]}%
\providecommand \BibitemOpen [0]{}%
\providecommand \bibitemStop [0]{}%
\providecommand \bibitemNoStop [0]{.\EOS\space}%
\providecommand \EOS [0]{\spacefactor3000\relax}%
\providecommand \BibitemShut  [1]{\csname bibitem#1\endcsname}%
\let\auto@bib@innerbib\@empty
\bibitem [{\citenamefont {Pyle}(1963)}]{Pyle1963}%
  \BibitemOpen
  \bibfield  {author} {\bibinfo {author} {\bibfnamefont {R.~W.}\ \bibnamefont
  {Pyle}},\ }\emph {\bibinfo {title} {Solid Torsional Horns}},\ \href
  {http://www.dtic.mil/cgi-bin/GetTRDoc?Location=U2&doc=GetTRDoc.pdf&AD=AD0408983}
  {Ph.D. thesis},\ \bibinfo  {school} {Harvard University, Cambridge} (\bibinfo
  {year} {1963})\BibitemShut {NoStop}%
\bibitem [{\citenamefont {Engan}\ \emph {et~al.}(1988)\citenamefont {Engan},
  \citenamefont {Byong},\ and\ \citenamefont {Blake}}]{Engan1988}%
  \BibitemOpen
  \bibfield  {author} {\bibinfo {author} {\bibfnamefont {H.}~\bibnamefont
  {Engan}}, \bibinfo {author} {\bibfnamefont {Y.}~\bibnamefont {Byong}}, \ and\
  \bibinfo {author} {\bibfnamefont {J.}~\bibnamefont {Blake}},\ }\href
  {\doibase 10.1109/50.4020} {\bibfield  {journal} {\bibinfo  {journal} {J.
  Lightwave Technol.}\ }\textbf {\bibinfo {volume} {6}},\ \bibinfo {pages}
  {428} (\bibinfo {year} {1988})}\BibitemShut {NoStop}%
\bibitem [{\citenamefont {Spinner}(1962)}]{Spinner1962}%
  \BibitemOpen
  \bibfield  {author} {\bibinfo {author} {\bibfnamefont {S.}~\bibnamefont
  {Spinner}},\ }\href@noop {} {\bibfield  {journal} {\bibinfo  {journal} {J.
  Am. Ceram. Soc.}\ }\textbf {\bibinfo {volume} {45}},\ \bibinfo {pages} {394}
  (\bibinfo {year} {1962})}\BibitemShut {NoStop}%
\bibitem [{\citenamefont {Borrelli}\ and\ \citenamefont
  {Miller}(1968)}]{Borrelli1968}%
  \BibitemOpen
  \bibfield  {author} {\bibinfo {author} {\bibfnamefont {N.~F.}\ \bibnamefont
  {Borrelli}}\ and\ \bibinfo {author} {\bibfnamefont {R.~A.}\ \bibnamefont
  {Miller}},\ }\href {\doibase 10.1364/AO.7.000745} {\bibfield  {journal}
  {\bibinfo  {journal} {Appl. Opt.}\ }\textbf {\bibinfo {volume} {7}},\
  \bibinfo {pages} {745} (\bibinfo {year} {1968})}\BibitemShut {NoStop}%
\bibitem [{\citenamefont {Pfaender}\ and\ \citenamefont
  {Schr{\"o}der}(1983)}]{Pfaender1983}%
  \BibitemOpen
  \bibfield  {author} {\bibinfo {author} {\bibfnamefont {H.~G.}\ \bibnamefont
  {Pfaender}}\ and\ \bibinfo {author} {\bibfnamefont {H.}~\bibnamefont
  {Schr{\"o}der}},\ }\href@noop {} {\emph {\bibinfo {title}
  {Schott-Glaslexikon}}}\ (\bibinfo  {publisher} {Moderne Verlags GmbH},\
  \bibinfo {year} {1983})\BibitemShut {NoStop}%
\bibitem [{\citenamefont {Chi}\ \emph {et~al.}(1984)\citenamefont {Chi},
  \citenamefont {Dennis~Jr.},\ and\ \citenamefont {Vossoughi}}]{Chi1984}%
  \BibitemOpen
  \bibfield  {author} {\bibinfo {author} {\bibfnamefont {M.}~\bibnamefont
  {Chi}}, \bibinfo {author} {\bibfnamefont {B.~G.}\ \bibnamefont {Dennis~Jr.}},
  \ and\ \bibinfo {author} {\bibfnamefont {J.}~\bibnamefont {Vossoughi}},\
  }\href {\doibase 10.1016/0022-460X(84)90581-9} {\bibfield  {journal}
  {\bibinfo  {journal} {J. Sound Vib.}\ }\textbf {\bibinfo {volume} {96}},\
  \bibinfo {pages} {235 } (\bibinfo {year} {1984})}\BibitemShut {NoStop}%
\bibitem [{\citenamefont {Brambilla}\ \emph {et~al.}(2009)\citenamefont
  {Brambilla}, \citenamefont {Xu}, \citenamefont {Horak}, \citenamefont {Jung},
  \citenamefont {Koizumi}, \citenamefont {Sessions}, \citenamefont
  {Koukharenko}, \citenamefont {Feng}, \citenamefont {Murugan}, \citenamefont
  {Wilkinson},\ and\ \citenamefont {Richardson}}]{Brambilla2009}%
  \BibitemOpen
  \bibfield  {author} {\bibinfo {author} {\bibfnamefont {G.}~\bibnamefont
  {Brambilla}}, \bibinfo {author} {\bibfnamefont {F.}~\bibnamefont {Xu}},
  \bibinfo {author} {\bibfnamefont {P.}~\bibnamefont {Horak}}, \bibinfo
  {author} {\bibfnamefont {Y.}~\bibnamefont {Jung}}, \bibinfo {author}
  {\bibfnamefont {F.}~\bibnamefont {Koizumi}}, \bibinfo {author} {\bibfnamefont
  {N.~P.}\ \bibnamefont {Sessions}}, \bibinfo {author} {\bibfnamefont
  {E.}~\bibnamefont {Koukharenko}}, \bibinfo {author} {\bibfnamefont
  {X.}~\bibnamefont {Feng}}, \bibinfo {author} {\bibfnamefont {G.~S.}\
  \bibnamefont {Murugan}}, \bibinfo {author} {\bibfnamefont {J.~S.}\
  \bibnamefont {Wilkinson}}, \ and\ \bibinfo {author} {\bibfnamefont {D.~J.}\
  \bibnamefont {Richardson}},\ }\href {\doibase 10.1364/AOP.1.000107}
  {\bibfield  {journal} {\bibinfo  {journal} {Adv. Opt. Photon.}\ }\textbf
  {\bibinfo {volume} {1}},\ \bibinfo {pages} {107} (\bibinfo {year}
  {2009})}\BibitemShut {NoStop}%
\bibitem [{\citenamefont {Vetsch}\ \emph {et~al.}(2010)\citenamefont {Vetsch},
  \citenamefont {Reitz}, \citenamefont {Sagu\'e}, \citenamefont {Schmidt},
  \citenamefont {Dawkins},\ and\ \citenamefont {Rauschenbeutel}}]{Vetsch2010}%
  \BibitemOpen
  \bibfield  {author} {\bibinfo {author} {\bibfnamefont {E.}~\bibnamefont
  {Vetsch}}, \bibinfo {author} {\bibfnamefont {D.}~\bibnamefont {Reitz}},
  \bibinfo {author} {\bibfnamefont {G.}~\bibnamefont {Sagu\'e}}, \bibinfo
  {author} {\bibfnamefont {R.}~\bibnamefont {Schmidt}}, \bibinfo {author}
  {\bibfnamefont {S.~T.}\ \bibnamefont {Dawkins}}, \ and\ \bibinfo {author}
  {\bibfnamefont {A.}~\bibnamefont {Rauschenbeutel}},\ }\href {\doibase
  10.1103/PhysRevLett.104.203603} {\bibfield  {journal} {\bibinfo  {journal}
  {Phys. Rev. Lett.}\ }\textbf {\bibinfo {volume} {104}},\ \bibinfo {pages}
  {203603} (\bibinfo {year} {2010})}\BibitemShut {NoStop}%
\bibitem [{\citenamefont {Goban}\ \emph {et~al.}(2012)\citenamefont {Goban},
  \citenamefont {Choi}, \citenamefont {Alton}, \citenamefont {Ding},
  \citenamefont {Lacro\^ute}, \citenamefont {Pototschnig}, \citenamefont
  {Thiele}, \citenamefont {Stern},\ and\ \citenamefont {Kimble}}]{Goban2012}%
  \BibitemOpen
  \bibfield  {author} {\bibinfo {author} {\bibfnamefont {A.}~\bibnamefont
  {Goban}}, \bibinfo {author} {\bibfnamefont {K.~S.}\ \bibnamefont {Choi}},
  \bibinfo {author} {\bibfnamefont {D.~J.}\ \bibnamefont {Alton}}, \bibinfo
  {author} {\bibfnamefont {D.}~\bibnamefont {Ding}}, \bibinfo {author}
  {\bibfnamefont {C.}~\bibnamefont {Lacro\^ute}}, \bibinfo {author}
  {\bibfnamefont {M.}~\bibnamefont {Pototschnig}}, \bibinfo {author}
  {\bibfnamefont {T.}~\bibnamefont {Thiele}}, \bibinfo {author} {\bibfnamefont
  {N.~P.}\ \bibnamefont {Stern}}, \ and\ \bibinfo {author} {\bibfnamefont
  {H.~J.}\ \bibnamefont {Kimble}},\ }\href {\doibase
  10.1103/PhysRevLett.109.033603} {\bibfield  {journal} {\bibinfo  {journal}
  {Phys. Rev. Lett.}\ }\textbf {\bibinfo {volume} {109}},\ \bibinfo {pages}
  {033603} (\bibinfo {year} {2012})}\BibitemShut {NoStop}%
\bibitem [{\citenamefont {Wuttke}\ and\ \citenamefont
  {Rauschenbeutel}(2013)}]{Wuttke2013}%
  \BibitemOpen
  \bibfield  {author} {\bibinfo {author} {\bibfnamefont {C.}~\bibnamefont
  {Wuttke}}\ and\ \bibinfo {author} {\bibfnamefont {A.}~\bibnamefont
  {Rauschenbeutel}},\ }\href {\doibase 10.1103/PhysRevLett.111.024301}
  {\bibfield  {journal} {\bibinfo  {journal} {Phys. Rev. Lett.}\ }\textbf
  {\bibinfo {volume} {111}},\ \bibinfo {pages} {024301} (\bibinfo {year}
  {2013})}\BibitemShut {NoStop}%
\bibitem [{\citenamefont {Warken}\ \emph {et~al.}(2008)\citenamefont {Warken},
  \citenamefont {Rauschenbeutel},\ and\ \citenamefont
  {Bartholom\"aus}}]{Warken2008}%
  \BibitemOpen
  \bibfield  {author} {\bibinfo {author} {\bibfnamefont {F.}~\bibnamefont
  {Warken}}, \bibinfo {author} {\bibfnamefont {A.}~\bibnamefont
  {Rauschenbeutel}}, \ and\ \bibinfo {author} {\bibfnamefont {T.}~\bibnamefont
  {Bartholom\"aus}},\ }\href@noop {} {\bibfield  {journal} {\bibinfo  {journal}
  {Photon. Spectra}\ }\textbf {\bibinfo {volume} {42}},\ \bibinfo {pages} {73}
  (\bibinfo {year} {2008})}\BibitemShut {NoStop}%
\bibitem [{\citenamefont {Stiebeiner}\ \emph {et~al.}(2010)\citenamefont
  {Stiebeiner}, \citenamefont {Garcia-Fernandez},\ and\ \citenamefont
  {Rauschenbeutel}}]{Stiebeiner2010}%
  \BibitemOpen
  \bibfield  {author} {\bibinfo {author} {\bibfnamefont {A.}~\bibnamefont
  {Stiebeiner}}, \bibinfo {author} {\bibfnamefont {R.}~\bibnamefont
  {Garcia-Fernandez}}, \ and\ \bibinfo {author} {\bibfnamefont
  {A.}~\bibnamefont {Rauschenbeutel}},\ }\href {\doibase 10.1364/OE.18.022677}
  {\bibfield  {journal} {\bibinfo  {journal} {Opt. Express}\ }\textbf {\bibinfo
  {volume} {18}},\ \bibinfo {pages} {22677} (\bibinfo {year}
  {2010})}\BibitemShut {NoStop}%
\bibitem [{\citenamefont {Wiedemann}\ \emph {et~al.}(2010)\citenamefont
  {Wiedemann}, \citenamefont {Karapetyan}, \citenamefont {Dan}, \citenamefont
  {Pritzkau}, \citenamefont {Alt}, \citenamefont {Irsen},\ and\ \citenamefont
  {Meschede}}]{Wiedemann2010}%
  \BibitemOpen
  \bibfield  {author} {\bibinfo {author} {\bibfnamefont {U.}~\bibnamefont
  {Wiedemann}}, \bibinfo {author} {\bibfnamefont {K.}~\bibnamefont
  {Karapetyan}}, \bibinfo {author} {\bibfnamefont {C.}~\bibnamefont {Dan}},
  \bibinfo {author} {\bibfnamefont {D.}~\bibnamefont {Pritzkau}}, \bibinfo
  {author} {\bibfnamefont {W.}~\bibnamefont {Alt}}, \bibinfo {author}
  {\bibfnamefont {S.}~\bibnamefont {Irsen}}, \ and\ \bibinfo {author}
  {\bibfnamefont {D.}~\bibnamefont {Meschede}},\ }\href {\doibase
  10.1364/OE.18.007693} {\bibfield  {journal} {\bibinfo  {journal} {Opt.
  Express}\ }\textbf {\bibinfo {volume} {18}},\ \bibinfo {pages} {7693}
  (\bibinfo {year} {2010})}\BibitemShut {NoStop}%
\bibitem [{\citenamefont {Vedam}(1950)}]{Vedam1950}%
  \BibitemOpen
  \bibfield  {author} {\bibinfo {author} {\bibfnamefont {K.}~\bibnamefont
  {Vedam}},\ }\href {\doibase 10.1103/PhysRev.78.472.2} {\bibfield  {journal}
  {\bibinfo  {journal} {Phys. Rev.}\ }\textbf {\bibinfo {volume} {78}},\
  \bibinfo {pages} {472} (\bibinfo {year} {1950})}\BibitemShut {NoStop}%
\bibitem [{\citenamefont {Wuttke}\ \emph {et~al.}(2012)\citenamefont {Wuttke},
  \citenamefont {Becker}, \citenamefont {Br\"uckner}, \citenamefont
  {Rothhardt},\ and\ \citenamefont {Rauschenbeutel}}]{Wuttke2012}%
  \BibitemOpen
  \bibfield  {author} {\bibinfo {author} {\bibfnamefont {C.}~\bibnamefont
  {Wuttke}}, \bibinfo {author} {\bibfnamefont {M.}~\bibnamefont {Becker}},
  \bibinfo {author} {\bibfnamefont {S.}~\bibnamefont {Br\"uckner}}, \bibinfo
  {author} {\bibfnamefont {M.}~\bibnamefont {Rothhardt}}, \ and\ \bibinfo
  {author} {\bibfnamefont {A.}~\bibnamefont {Rauschenbeutel}},\ }\href@noop {}
  {\bibfield  {journal} {\bibinfo  {journal} {Opt. Lett.}\ }\textbf {\bibinfo
  {volume} {37}} (\bibinfo {year} {2012})}\BibitemShut {NoStop}%
\bibitem [{\citenamefont {Pohl}\ \emph {et~al.}(2002)\citenamefont {Pohl},
  \citenamefont {Liu},\ and\ \citenamefont {Thompson}}]{Pohl2002}%
  \BibitemOpen
  \bibfield  {author} {\bibinfo {author} {\bibfnamefont {R.~O.}\ \bibnamefont
  {Pohl}}, \bibinfo {author} {\bibfnamefont {X.}~\bibnamefont {Liu}}, \ and\
  \bibinfo {author} {\bibfnamefont {E.}~\bibnamefont {Thompson}},\ }\href
  {\doibase 10.1103/RevModPhys.74.991} {\bibfield  {journal} {\bibinfo
  {journal} {Rev. Mod. Phys.}\ }\textbf {\bibinfo {volume} {74}},\ \bibinfo
  {pages} {991} (\bibinfo {year} {2002})}\BibitemShut {NoStop}%
\bibitem [{\citenamefont {Penn}\ \emph {et~al.}(2006)\citenamefont {Penn},
  \citenamefont {Ageev}, \citenamefont {Busby}, \citenamefont {Harry},
  \citenamefont {Gretarsson}, \citenamefont {Numata},\ and\ \citenamefont
  {Willems}}]{Penn2006}%
  \BibitemOpen
  \bibfield  {author} {\bibinfo {author} {\bibfnamefont {S.~D.}\ \bibnamefont
  {Penn}}, \bibinfo {author} {\bibfnamefont {A.}~\bibnamefont {Ageev}},
  \bibinfo {author} {\bibfnamefont {D.}~\bibnamefont {Busby}}, \bibinfo
  {author} {\bibfnamefont {G.~M.}\ \bibnamefont {Harry}}, \bibinfo {author}
  {\bibfnamefont {A.~M.}\ \bibnamefont {Gretarsson}}, \bibinfo {author}
  {\bibfnamefont {K.}~\bibnamefont {Numata}}, \ and\ \bibinfo {author}
  {\bibfnamefont {P.}~\bibnamefont {Willems}},\ }\href {\doibase
  10.1016/j.physleta.2005.11.046} {\bibfield  {journal} {\bibinfo  {journal}
  {Phys. Lett. A}\ }\textbf {\bibinfo {volume} {352}},\ \bibinfo {pages} {3}
  (\bibinfo {year} {2006})}\BibitemShut {NoStop}%
\bibitem [{\citenamefont {Spillane}\ \emph {et~al.}(2003)\citenamefont
  {Spillane}, \citenamefont {Kippenberg}, \citenamefont {Painter},\ and\
  \citenamefont {Vahala}}]{Spillane2003}%
  \BibitemOpen
  \bibfield  {author} {\bibinfo {author} {\bibfnamefont {S.~M.}\ \bibnamefont
  {Spillane}}, \bibinfo {author} {\bibfnamefont {T.~J.}\ \bibnamefont
  {Kippenberg}}, \bibinfo {author} {\bibfnamefont {O.~J.}\ \bibnamefont
  {Painter}}, \ and\ \bibinfo {author} {\bibfnamefont {K.~J.}\ \bibnamefont
  {Vahala}},\ }\href {\doibase 10.1103/PhysRevLett.91.043902} {\bibfield
  {journal} {\bibinfo  {journal} {Phys. Rev. Lett.}\ }\textbf {\bibinfo
  {volume} {91}},\ \bibinfo {pages} {043902} (\bibinfo {year}
  {2003})}\BibitemShut {NoStop}%
\bibitem [{\citenamefont {Demtr\"oder}(2003)}]{Demtroeder1}%
  \BibitemOpen
  \bibfield  {author} {\bibinfo {author} {\bibfnamefont {W.}~\bibnamefont
  {Demtr\"oder}},\ }\href@noop {} {\emph {\bibinfo {title} {Experimentalphysik
  1}}},\ \bibinfo {edition} {3rd}\ ed.,\ Vol.~\bibinfo {volume} {1}\ (\bibinfo
  {publisher} {Springer Berlin / Heidelberg},\ \bibinfo {year}
  {2003})\BibitemShut {NoStop}%
\end{thebibliography}
%

\end{document}